\documentclass[journal]{IEEEtran}
\ifCLASSINFOpdf
   \usepackage[pdftex]{graphicx}
   \graphicspath{{../pdf/}{../jpeg/}}
   \DeclareGraphicsExtensions{.pdf,.jpeg,.png}
\else
   \usepackage[dvips]{graphicx}
   \usepackage{tabularx}
   \graphicspath{{../eps/}}
   \DeclareGraphicsExtensions{.eps}
\fi
\usepackage{epstopdf}
\usepackage{amsmath}
\usepackage{bm}
\usepackage{amsthm}
\UseRawInputEncoding
\usepackage{subfigure}
\usepackage{amssymb}
\usepackage{cite}
\usepackage{color}
\usepackage{upgreek}
\begin{document}
\title{\huge An Experimental Validation of Reconfigurable Intelligent Surfaces Achieving Pulse Width-Modulated Singular Reflection Angles Without External Power Sources}
\author{Eisuke~Omori, Kairi~Takimoto, Atsuko~Nagata, Ashif~Fathnan~\IEEEmembership{Member,~IEEE,}
        Shinya~Sugiura,~\IEEEmembership{Senior Member,~IEEE,}
        and~Hiroki~Wakatsuchi,~\IEEEmembership{Member,~IEEE}
\thanks{E. Omori, K. Takimoto, A. Nagata, A. Fathnan, and H. Wakatsuchi are with the Department
of Engineering, Nagoya Institute of Technology, Aichi, 466-8555 Japan (e-mail: wakatsuchi.hiroki@nitech.ac.jp).}
\thanks{S. Sugiura is with the Institute of Industrial Science, The University of Tokyo, Tokyo, 153-8505, Japan.}
\thanks{Manuscript received January 31, 2024. This work was supported in part by the Japan Science and Technology Agency (JST) under the Precursory Research for Embryonic Science and Technology (PRESTO) Program (nos.\ JPMJPR1933 and JPMJPR193A), Fusion Oriented Research for Disruptive Science and Technology (no.\ JPMJFR222T), and Adopting Sustainable Partnerships for Innovative Research Ecosystem (ASPIRE) (nos.\ JPMJAP2345 and JPMJAP2431); KAKENHI grants from the Japan Society for the Promotion of Science (JSPS) (nos.\ 21H01324, 22F22359, and 23H00470); and the National Institute of Information and Communications Technology (NICT), Japan, under commissioned research (no.\ JPJ012368C06201).}
}

\makeatletter
\def\ps@IEEEtitlepagestyle{
  \def\@oddfoot{\mycopyrightnotice}
  \def\@evenfoot{}
}
\def\mycopyrightnotice{
  {\footnotesize
  \begin{minipage}{\textwidth}
  \centering
  Copyright~\copyright~2023 IEEE. Personal use of this material is permitted. However, permission to use this  \\ 
  material for any other purposes must be obtained from the IEEE by sending a request to pubs-permissions@ieee.org.
  \end{minipage}
  }
}

\maketitle

\begin{abstract}
In this study, we introduce a design concept that leverages pulse width variation to enable a reconfigurable intelligent surface (RIS) and to autonomously switch reflection properties between two angles without any active control system. Our RIS alters its beam pattern from a singular specular reflection to another unique singular anomalous reflection when the incoming waveform changes from a short pulse to a continuous wave, even at the same frequency. Unlike conventional RISs, our passive control mechanism eliminates the requirements of active components and precise symbol-level synchronization with the transmitting antennas, reducing the system complexity level while offering dynamic material adaptability. We numerically show that the proposed RIS design is capable of varying the received magnitude of an incident wave by a factor of ten, which is also experimentally validated for the first time. Employing binary phase-shift keying (BPSK) modulation, we further report that the communication characteristics can be varied by 7 dB or more, which indicates that the proposed design is not limited to a single frequency component as long as the bandwidth of the given signal is covered by that of the RIS design. These results may present new opportunities for exploring and deploying pulse width-dependent RISs in practical scenarios involving next-generation communication systems.
\end{abstract}

\begin{IEEEkeywords}
\textbf{reconfigurable intelligent surfaces, metasurface, nonlinear embedded circuits}
\end{IEEEkeywords}

\section{Introduction}

\IEEEPARstart{S}{mart} radio environments have recently been developed for wireless communication systems to address diverse challenges, such as increased data rates, seamless user mobility, and increased energy efficiency rates \cite{renzo2019smart, dang2020should, di2020smart, gradoni1smart}. These innovations are aided by reconfigurable intelligent surfaces (RISs), which include small engineered structures that are referred to as metasurfaces \cite{sievenpiper1999high, yu2011light, yu2014flat}. Through their subwavelength unit cells, these structures can artificially and independently control permittivity, permeability, refractive index, and wave impedance \cite{pfeiffer2013metamaterial, pendryENG1, pendryMNG, smithDNG1D, smithDNG2D2} so that RISs can be strategically designed to deliberately interact with the target wireless communication signals \cite{elmossallamy2020reconfigurable}. Utilizing these intelligent surfaces, dynamic control over the signal reflection and propagation processes can be attained, thereby integrating adaptability into the radio environment while simultaneously optimizing signal paths and mitigating issues such as signal blockage and interference \cite{CCemcBook}. By leveraging RIS technology, smart radio environments can enable more efficient, flexible, and responsive wireless networks, advancing communication solutions beyond the traditional limitations of the existing approaches \cite{wu2019towards, zhang2021wireless}.

\begin{figure}
  \begin{center}
  \includegraphics[width=\linewidth]{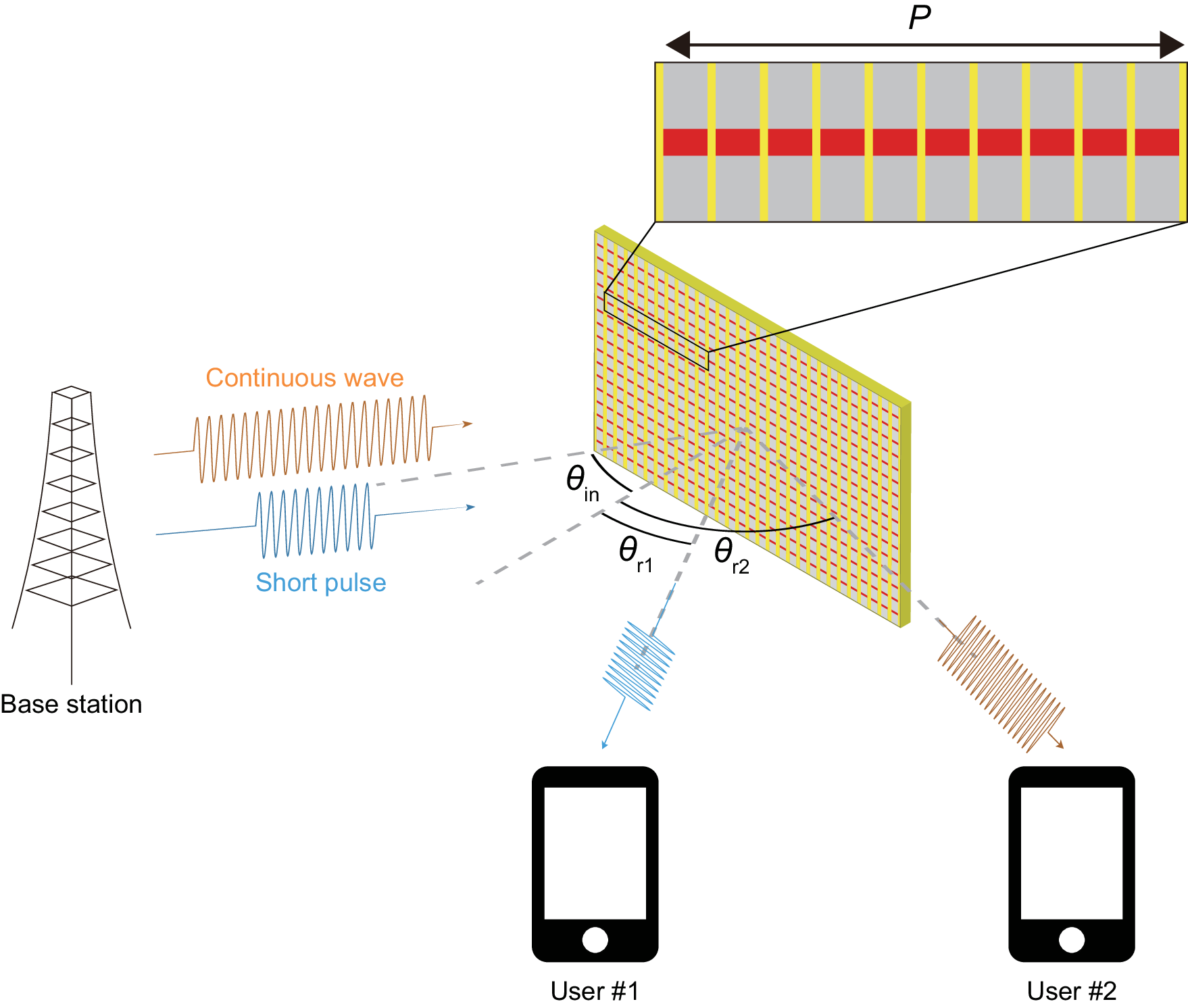}\\
  \caption{A visualization of the proposed RIS
 and its ability to manipulate the beamforming process based on the pulse widths of the incoming wireless signals without an external power source (e.g., a direct-current (DC) supply) as a passive solution to outdoor communications. The incident angle $\theta_{in}$ and reflected angles $\theta_{r1}$ and $\theta_{r2}$ can be customized by the RIS design. }\label{fig:1}
  \end{center}
\end{figure}

Various practical schemes, such as varactor-based reconfigurable unit cells \cite{araghi2022reconfigurable,pei2021ris}, 
PIN 
diode-based unit cells \cite{trichopoulos2022design}, and graphene meta-atoms \cite{molero2021metamaterial}, have been proposed for realizing such beamforming RISs. Despite these techniques having demonstrated diverse wave manipulation capabilities, the resulting RISs often demand dedicated controllers and intricate biasing for implementing state alterations. In conventional RIS-based systems, beam training is a prerequisite for computing RIS element control weights before transmitting data, necessitating synchronization between a transmitter and each RIS element \cite{zheng2022survey, ino2023noncoherent}. Achieving precise synchronization between a base station (BS) and a receiver is also challenging, especially in fast-changing environments with short coherence times, due to the heavy reliance of RISs on receiver feedback \cite{elmossallamy2020reconfigurable}. A pilot-based channel estimator with a small number of sensing devices can address the synchronization issues \cite{taha2021enabling} encountered in RIS systems, but this simultaneously introduces complexities and costs due to the need for multiple control lines between sensors and unit cells, particularly in setups with numerous elements.

In this study, to address the aforementioned challenges, we introduce an RIS featuring unit cells that autonomously adjust in response to incoming waveform characteristics, as depicted in Fig.\ \ref{fig:1}. Unlike conventional diode-based RISs that are reliant on external control for biasing, our unit cells are directly controlled by embedded circuits that are sensitive to the pulse widths of the incoming wireless signals. Consequently, reconfigurability is achieved through passive unit cell tuning, eliminating the need for active tuning mechanisms. This novel RIS design facilitates dynamic beamforming without the complexities associated with control lines and precise synchronization with the BS. To this end, the proposed RIS leverages subwavelength structures, which are referred to as waveform-selective metasurfaces \cite{wakatsuchi2013waveform,vellucci2019waveform,barbuto2020waveguide,f2020temporal,barbuto2021metasurfaces,homma2022anisotropic,fathnan2022method,ushikoshi2023pulse}. Our RIS offers the potential to selectively steer communication radio waves on the basis of pulse width control when integrated into a wireless environment. Specifically, advancing upon our prior study that addressed the reconfigurability between a diffraction grating and an anomalous reflection generating unavoidable multiple beams \cite{fathnan2023unsynchronized}, our current work achieves a beamforming concept that transitions from singular specular reflection to another singular anomalous reflection modality based on the incident waveform type. This system provides an advantage in terms of altering the communication configuration for a single receiver while ensuring high signal reception quality at two predefined locations (these features were absent from our previous design). Additionally, we present the first experimental findings that ensure the inclusion of numerical simulation results, demonstrating the feasibility of implementing our RIS concept in real-world applications.

The rest of this article is composed as follows. In Section II, we explain a fundamental beamforming concept along with the design approach and working mechanism of waveform-selective (i.e., pulse width-dependent) metasurfaces. These metasurface unit cells are assembled as supercells of RISs and numerically tested in Section III. In this section, first, we simplify the metasurface units as sheet impedances to readily estimate their reflecting profiles, which are more precisely evaluated with rigorous simulation models later. The proposed RIS design is experimentally validated in Section IV. We further evaluate the performance of the proposed RIS design from the viewpoint of its communication characteristics in Section V. These results and their implications are discussed in Section VI, which is followed by the conclusion in Section VII.

\section{Beamforming Concept and Realization of Unit Cells}

\begin{figure}
\subfigure[]{
\centering
\includegraphics[width=0.58\linewidth]{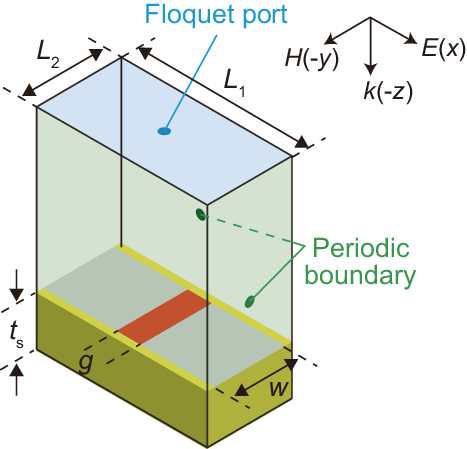}
}
\subfigure[]{
\centering
\includegraphics[width=0.35\linewidth]{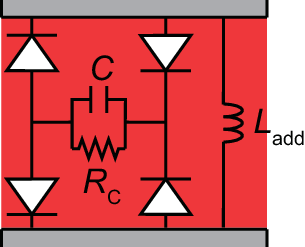}
}
\caption{The periodic unit cell configuration. (a) A unit cell configuration consisting of metallic patches (gray) on top of a grounded dielectric substrate (yellow). (b) A circuit configuration with a connection between the metallic patches. The detailed geometrical and circuit parameters are shown in Tables \ref{tab:1} to \ref{tab:3}. }
\label{fig:2}
\end{figure}

\begin{table} 
\caption{The geometric parameters adopted in the fundamental unit cell shown in Fig.\ \ref{fig:2}. All of the numbers are in mm. \label{tab:1}}
\renewcommand{\arraystretch}{1.2}
\centering
\begin{tabular}{|c| c|}
\hline
Parameter                   & Value \\ \hline\hline
$t_s$                            & 3.04                              \\ 
$g$                         & 1                                         \\ 
$w$                            & 16                      \\ 
$L_1$                           & 34    \\ 
$L_2$                            & 17                     \\ 
\hline
\end{tabular}
\vspace{5mm}
%
\caption{The parameters of the discrete circuit components that are adopted in the fundamental unit cell of Fig.\ \ref{fig:2}. $L_{add}$ is fixed at 4 nH in Fig.\ \ref{fig:2b} while it is varied in Fig.\ \ref{fig:3} as summarized in Table \ref{tab:4}. \label{tab:2}}
\renewcommand{\arraystretch}{1.2}
\centering
\begin{tabular}{|c| c| c|}
\hline
Parameter & Unit                  & Value \\ \hline\hline
$C$        & nF                     & 1                              \\ 
$R_C$      & k$\Omega$                    & 10                                        \\ 
\hline
\end{tabular}
\vspace{5mm}
%
\caption{The SPICE parameters of the diodes used in Fig.\ \ref{fig:2}. \label{tab:3}}
\renewcommand{\arraystretch}{1.2}
\centering
\begin{tabular}{|c| c| c|}
\hline
Parameter & Unit                  & Value \\ \hline\hline
$B_V$        & V                     & 7.0                              \\ 
$C_{j0}$      & pF                    & 0.18                                         \\ 
$E_G$        & eV                    & 0.69                      \\ 
$I_{BV}$       & A                     & 1 $\times$ 10$^{-5}$    \\ 
$I_S$        & A                     & 5 $\times$ 10$^{-8}$                     \\ 
$N$         &                       & 1.08                        \\ 
$R_S$        & $\Omega$ & 6.0                      \\ 
$P_B$ (VJ)    & V                     & 0.65                    \\ 
$P_T$ (XTI)   &                       & 2                        \\ 
$M$         &                       & 0.5                      \\ \hline
\end{tabular}
\end{table}

The proposed RIS employs beamforming methods to alter the directions of incoming radio waves via passive adjustments with its unit cells on the basis of the configuration shown in Fig.\ \ref{fig:2} and its design parameters, which are summarized in Tables \ref{tab:1} to \ref{tab:3}. This setup includes subwavelength reflective components that are constructed using a metal--insulator--metal (MIM) configuration, as shown in Fig.\ \ref{fig:2}(a). In this design, a dielectric layer (Rogers RO3003) is positioned between a fully metallic ground plane (below) and a patterned metallic layer (above). This configuration has been employed in many other studies \cite{sievenpiper1999high, engheta2006metamaterials}, as well as in our earlier studies in which the same unit cell arrangement was utilized \cite{wakatsuchi2013waveform, takimoto2023inkjet, cheng2023TAP}. However, in our earlier studies, we exploited unit cell switching between two states, including not only a phase-variant state that was suitable for anomalous reflection but also an absorbing state that was unsuitable for efficient reflection. Instead of the absorbing state, the operating unit cell is extended in this study to include a third state that represents a normal specular reflection, as explained below. First, the use of conducting patches above the ground leads to the design of an artificial magnetic conductor, which produces a 0-degree reflection phase \cite{sievenpiper1999high}. This condition is readily accounted for by 
\begin{eqnarray}
   R=\frac{Z-Z_0}{Z-Z_0}
   \label{eq:1}
\end{eqnarray}
and
\begin{eqnarray}
   Z=\Biggl(\frac{1}{R_0}+\frac{1}{j\omega L_0}+j\omega C_0\Biggr)^{-1},
   \label{eq:2}
\end{eqnarray}
where $R$, $Z$, $Z_0$, $R_0$, $L_0$, and $C_0$ denote the complex reflection coefficient; the wave impedance of the metasurface unit cell; the wave impedance in a vacuum; and the entire resistive, inductive, and capacitive components of the metasurface unit cell, respectively. Additionally, $j^2=-1$, and $\omega$ represents the angular frequency (i.e., $\omega=2\pi f$, where $f$ is the frequency). Here, $C_0$ is mostly determined by the patch geometry and conductor gap; specifically, \cite{sievenpiperPhD}
\begin{eqnarray}
   C_0=\frac{(L_1-g)(\epsilon_0(1+\epsilon_r))}{\pi}\cosh^{-1}{\frac{L_1}{g}},
   \label{eq:3}
\end{eqnarray}
where $\epsilon_0$ and $\epsilon_r$ are the permittivity of a vacuum and the relative permittivity of the substrate ($\epsilon_r$ = 3.0 in this study), respectively. In contrast, assuming that the substrate is a nonmagnetic medium, $L_0$ is related to the substrate thickness \cite{luukkonen2009thin}:
\begin{eqnarray}
   L_0=\mu_0t_s,
   \label{eq:4}
\end{eqnarray}
where $\mu_0$ is the permeability of a vacuum. Therefore, by properly adjusting these physical dimensions, the reflection phase can be set not only to 0 degrees but also to other values \cite{sievenpiper2003two}. We further note that the conductor design can be customized to a totally different geometry for narrowband, broadband, or multiband operations. 

\begin{figure}
\subfigure[]{
\centering
\includegraphics[width=\linewidth]{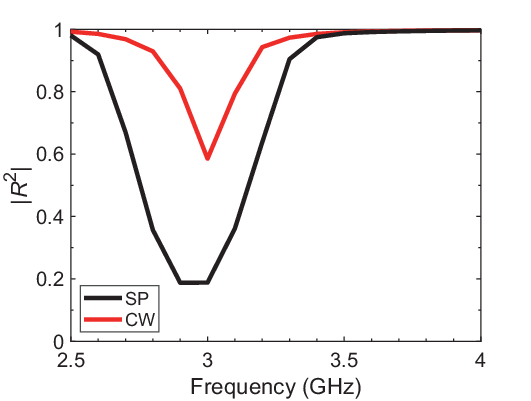}
}
\subfigure[]{
\centering
\includegraphics[width=\linewidth]{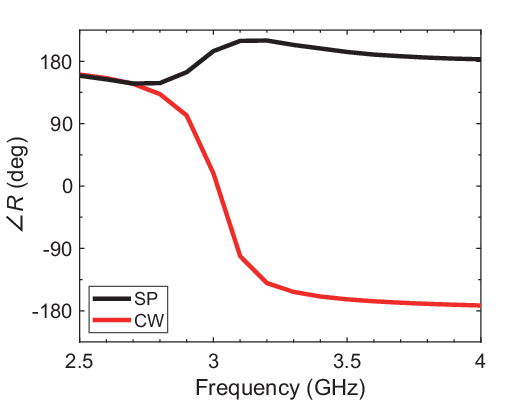}
}
\caption{The reflecting profiles of the periodic unit cell shown in Fig.\ \ref{fig:2}. (a) The reflection magnitudes and (b) phases for the 50-ns SPs and CWs.}
\label{fig:2b}
\end{figure}

More importantly, this fundamental reflection phase design attains a higher degree of freedom in our unit cell drawn in Fig.\ \ref{fig:2}(b) by introducing a set of four diodes (Avago, HSMS 286x series) between the conductor edges. These diodes play the role of a diode bridge, thereby fully rectifying the electric charges induced by the incoming wave and generating an infinite set of frequency components \cite{wakatsuchi2013waveform}. Importantly, however, the largest frequency component appears at a frequency of zero, as expected from the Fourier series expansion of $|\sin|$. Therefore, by including a reactive component and a resistive component within the diode bridge, we can exploit the time-varying responses associated with the transients of classic direct-current (DC) circuits. These circuit responses are further coupled to the electromagnetic responses of the metasurface in the time domain. More specifically, a capacitor $C$ is connected to a resistor $R_C$ in parallel within the diode bridge to fully store the energy of an incoming short pulse (SP). The stored energy is then dissipated by $R_C$. On the other hand, when a sufficiently long pulse or a continuous wave (CW) comes in, $C$ is fully charged so that the induced electric charges are no longer allowed to enter the diode bridge. We also add an additional inductor $L_{add}$ to control the reflection phase. By using a cosimulation method that is available in ANSYS Electronics Desktop (2023 R2) and the SPICE parameters of Table \ref{tab:3}, the pulse width-dependent reflecting profile of our metasurface is obtained as a function of the frequency in Fig.\ \ref{fig:2b}, where the input power is set to 0 dBm. Here, we calculate the reflecting profiles of 50-ns SPs by dividing the entire reflecting energy by the entire incident energy in transient simulations. For the reflecting profiles of CWs, harmonic balance (HB) simulations are adopted to readily obtain the steady-state responses of our metasurface. Also, $L_{add}$ is set to 4 nH in Fig.\ \ref{fig:2b}. Under these circumstances, the reflection magnitudes observed for the SPs at approximately 2.9 GHz are lower than those obtained for CWs, which can be explained by the abovementioned pulse width-dependent mechanism. More importantly, the reflection phases of the SPs and CWs are largely different. Specifically, the CW condition exhibits a smooth transition between $\pm$180 degrees with increasing frequency because the capacitor is fully charged; thus, the diode bridge remains an open circuit and permits a gradient phase change. In contrast, the reflection phase is almost 180 degrees under the SP condition. In this case, the state of the diode bridge becomes a short circuit, which does not permit the gradual phase change observed under the CW condition. Note that the absorbing performance achieved for the CWs is relatively enhanced in Fig.\ \ref{fig:2b} solely to introduce the pulse width-dependent response of our metasurface. However, this absorptance magnitude is reduced in the subsequent part of this study to realize efficient reflection performance for RISs while varying or fixing their reflection phases in accordance with the incoming pulse width. 

\begin{figure}
    \centering
    \includegraphics[width=\linewidth]{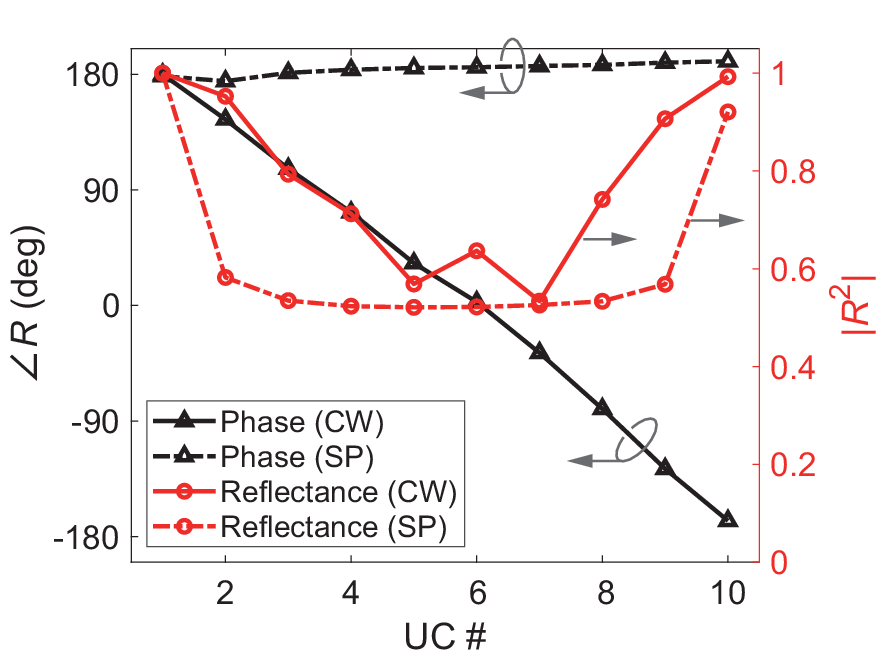}
    \caption{The reflection profiles produced for the unit cells at 3 GHz with various $L_{add}$ values (see Table \ref{tab:4} for the specific inductance values). }
    \label{fig:3}
\end{figure}

On the basis of the simulation results presented in Fig.\ \ref{fig:2b}, the supercell of our RIS is designed to operate at a frequency of 3 GHz (i.e., $\lambda = 100$ mm, where $\lambda$ represents the wavelength of the incident wave) and incorporate ten unit cells per period (i.e., $P = 170$ mm, where $P$ represents the periodicity of the supercells). These ten unit cells, numbered UC$\#$1 to UC$\#$10, are designed with the different $L_{add}$ values shown in Table \ref{tab:4} ($L_{add1}$ to $L_{add10}$, respectively), which enables us to tailor the reflecting profiles, as shown in Fig.\ \ref{fig:3}. Here, when SPs are used, the phase profile remains at approximately 180 degrees over the whole set of unit cells (the dashed lines in Fig.\ \ref{fig:3}); thus, the overall configuration is expected to result in a specular reflection. However, when CWs are used instead, the metasurface yields distinct reflection profiles, significantly amplifying the reflectance with a phase profile that covers $2\pi$ over the entire set of unit cells (the solid lines in Fig.\ \ref{fig:3}). Note that these $L_{add}$ values are determined to satisfy \cite{yu2011light}
\begin{eqnarray}
   P=\frac{\lambda}{\sin{\theta_{r2}-\sin{\theta _{in}}}}
   \label{eq:5}
\end{eqnarray}
and
\begin{eqnarray}
   \angle R(y)=\frac{\omega}{c}y\sin{\theta_{r2}},
   \label{eq:6}
\end{eqnarray}
where $\theta_{in}$, $\theta_{r2}$, and $c$ denote the incident angle, the designed anomalous reflection angle, and the speed of light, respectively. In realistic RIS configurations, Eq.\ (\ref{eq:6}) is discretized in space as follows:
\begin{eqnarray}
   \angle R_n=\frac{\omega}{c}nL_2\sin{\theta_{r2}},
   \label{eq:7}
\end{eqnarray}
where $n$ represents the number of constituent unit cells, namely, 1 to 10. Thus, the reflected wavefront is artificially tilted with respect to the specular reflection angle $\theta_{r1}$ (see Fig.\ \ref{fig:1}). Note that the anomalous reflection angle $\theta_{r2}$ is set to 30 degrees in this study for an incident angle $\theta_{in}$ of 0 degrees (but when $\theta_{in}=$ 10 degrees, $\theta_{r2}=$ 50 degrees), although $\theta_{r2}$ can be adjusted to other angles in accordance with Eqs.\ (\ref{eq:5}) and (\ref{eq:7}). Moreover, unlike in past studies \cite{fathnan2023unsynchronized}, our unit cells achieve not only a flat reflection phase for one of the switching states but also a smooth phase gradient for another state, which is made possible by adjusting the reflectance magnitude. Therefore, when these unit cells are assembled as periodic supercells, a normal specular reflection and an anomalous reflection are expected to be obtained for an SP and a CW, respectively, even at the same frequency. 

\begin{table}   
\caption{The $L_{add}$ values. $L_{add1}$ to $L_{add10}$ are used for UC$\#$1 to UC$\#$10, respectively (see Fig.\ \ref{fig:3} and later Fig.\ \ref{fig:5}). \label{tab:4}}
\renewcommand{\arraystretch}{1.2}
\centering
\begin{tabular}{|c |c |c|}
\hline
UC$\#$ & Parameter & Value (nH)\\ \hline\hline
1 &$L_{\mathrm{add1}}$     & 0.1                  \\ 
2 & $L_{\mathrm{add2}}$     & 3.1                            \\ 
3 & $L_{\mathrm{add3}}$ & 3.7                            \\ 
4 & $L_{\mathrm{add4}}$     & 4.0                   \\ 
5 &  $L_{\mathrm{add5}}$     & 4.2                   \\ 
6 & $L_{\mathrm{add6}}$     & 4.3                         \\ 
7 & $L_{\mathrm{add7}}$     & 4.5                    \\ 
8 & $L_{\mathrm{add8}}$     & 4.7                         \\ 
9 & $L_{\mathrm{add9}}$     & 5.3                           \\ 
10 & $L_{\mathrm{add10}}$     & 40.0                             \\ \hline
\end{tabular}
\end{table}

\section{Numerical Reflection Performance of the RIS\label{sec:sim}}
\begin{figure}
\subfigure[]{
\centering
\includegraphics[width=\linewidth]{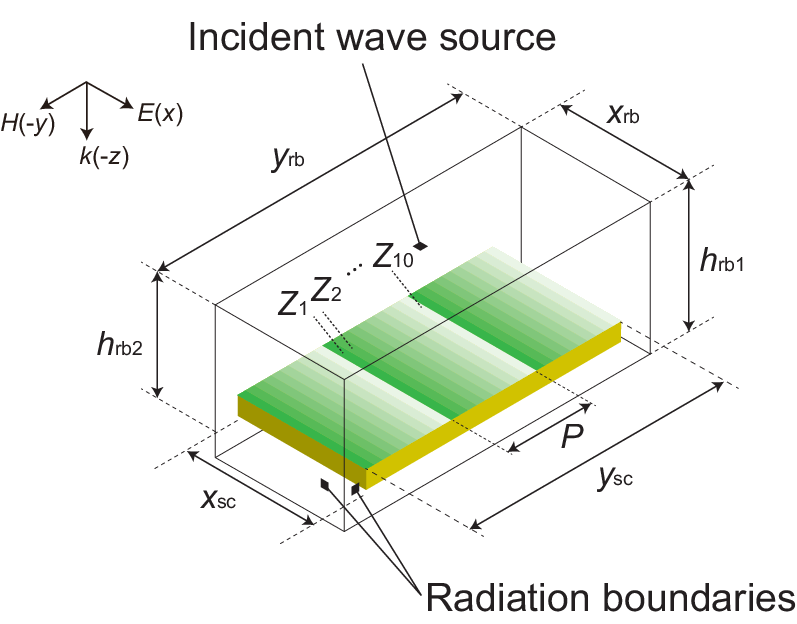}
}
\subfigure[]{
\centering
\includegraphics[width=\linewidth]{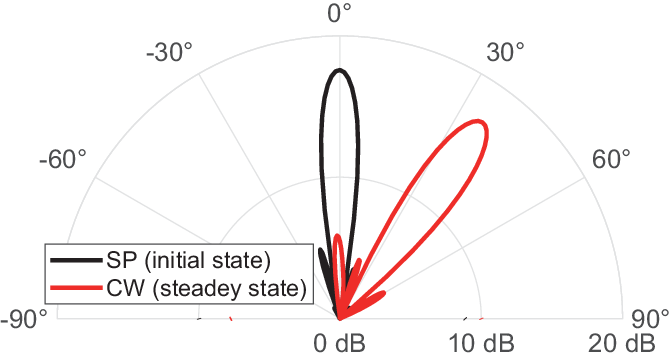}
}
\caption{(a). The simplified RIS model based on homogeneous impedance sheets representing the impedances of UC$\#$1 to UC$\#$10. The geometric parameters are given in Table \ref{tab:5}. The equivalent surface impedances, represented by $Z_1$ to $Z_{10}$, are shown in Table \ref{tab:6}. (b) The simulated reflection profiles at a normal incidence.}
\label{fig:4}
\end{figure}

Based on this configuration, we realized distinct beamforming profiles that were available from such a waveform-selective metasurface under two different incident wave illuminations. First, these reflecting profiles were validated via a simplified simulation approach. Here, we used the geometric parameters in Table \ref{tab:5} to model our RIS composed of three supercells, each of which had ten homogeneous impedance sheets representing the impedances of UC$\#$1 to UC$\#$10, as shown in Fig.\ \ref{fig:4}(a). These equivalent surface impedances, denoted by $Z_1$ to $Z_{10}$, were calculated as \cite{fathnan2023unsynchronized, holloway2005reflection}
\begin{eqnarray}
   Z_n=\frac{jR_n Z_s\tan{(\beta_st_s)}}{jZ_s\tan{(\beta_st_s)}-R_n},
   \label{eq:8}
\end{eqnarray}
where the subscripts $n$s are integers representing 1 to 10, while $Z_s$ and $\beta_s$ are the impedance and wavenumber of the substrates, respectively. These impedance sheets, characterized by their $Z_n$ values and summarized in Table \ref{tab:6}, were deployed on the substrate to readily emulate the complicated numerical analysis. Additionally, the boundary conditions were set as radiation boundaries, while the top boundary additionally illuminated an incident wave on the simplified RIS model at a normal angle. Under these circumstances, as shown in Fig.\ \ref{fig:4}(b), the incident wave under the SP condition was reflected at a normal angle as a specular reflection since the reflection phase of each impedance sheet was almost 180 degrees. Under these conditions, any sidelobe was at least 12.5 dB smaller than the normal reflection magnitude. However, when the surface impedance values were changed to those for the CW condition, the incident wave was primarily reflected at approximately 36.1 degrees as an anomalous reflection. The magnitude of the anomalous reflection was enhanced by approximately 19.2 dB over that of the reflection observed under the SP condition at the same angle (i.e., 36.1 degrees). Moreover, any sidelobe here was at least 11.2 dB smaller than the anomalous reflection peak. Additionally, note that the anomalous reflection peak under the CW condition was comparable to the specular reflection magnitude produced at 0 degrees under the SP condition, meaning that signals would be expected to be received at the same power level in both conditions. The reflection magnitudes attained at the designed angles were at least 11.6 dB larger than those achieved for a nondesigned angle (e.g., 0 degrees for the CW). On the basis of these results, the beamforming angle is expected to vary in accordance with the incident pulse width, even at a constant frequency. 

\begin{table} 
\caption{The geometric parameters adopted in the simplified RIS simulation model shown in Fig.\ \ref{fig:4}. All of the numbers are in mm. \label{tab:5}}
\renewcommand{\arraystretch}{1.2}
\centering
\begin{tabular}{|c| c|}
\hline
Parameter                   & Value \\ \hline\hline
$x_{sc}$                            & 340                              \\ 
$y_{sc}$                            & 680                                \\ 
$x_{rb}$                            & 340                              \\ 
$y_{rb}$                            & 830                                \\ 
$h_{rb1}$                            & 375                              \\ 
$h_{rb2}$                            & 300                                \\ 
\hline
\end{tabular}
\end{table}

\begin{table} 
\caption{The equivalent surface impedances adopted in the simplified RIS simulation model of Fig.\ \ref{fig:4}. \label{tab:6}}
\renewcommand{\arraystretch}{1.2}
\centering
\begin{tabular}{|c| c| c| c|}
\hline
UC$\#$                &Parameter   & Value for SPs ($\Omega$) & Value for CWs ($\Omega$)\\ \hline\hline
1                      &$Z_1$       &  -0.0152+j5.8885   & 0.0004+j0.9063                        \\ 
2                      &$Z_2$       &  41.6626-j50.2612     & 27.0480-j194.3000                       \\ 
3                      &$Z_3$        &  33.2234-j51.8956   & 8.3665-j99.7258                      \\ 
4                      &$Z_4$        &  31.0245-j52.2232    & 5.0703-j86.4987                     \\ 
5                      &$Z_5$      &    29.9751-j52.1277   & 4.8925-j78.6516                      \\ 
6                      &$Z_6$       &   29.5958-j51.9694   & 3.3114-j74.8931                       \\ 
7                      &$Z_7$      &   28.9364-j51.5067  & 4.3940-j70.2359                         \\ 
8                      &$Z_8$        &   28.4203-j50.8038 & 2.7466-j64.1845                        \\ 
9                      &$Z_9$        &   27.2000-j47.9575   & 1.8645-j53.4862                     \\ 
10                     &$Z_{10}$        & 7.3011-j24.2552   & 0.5763-j26.3412                         \\ 
\hline
\end{tabular}
\end{table}

\begin{figure}
    \centering
    \includegraphics[width=\linewidth]{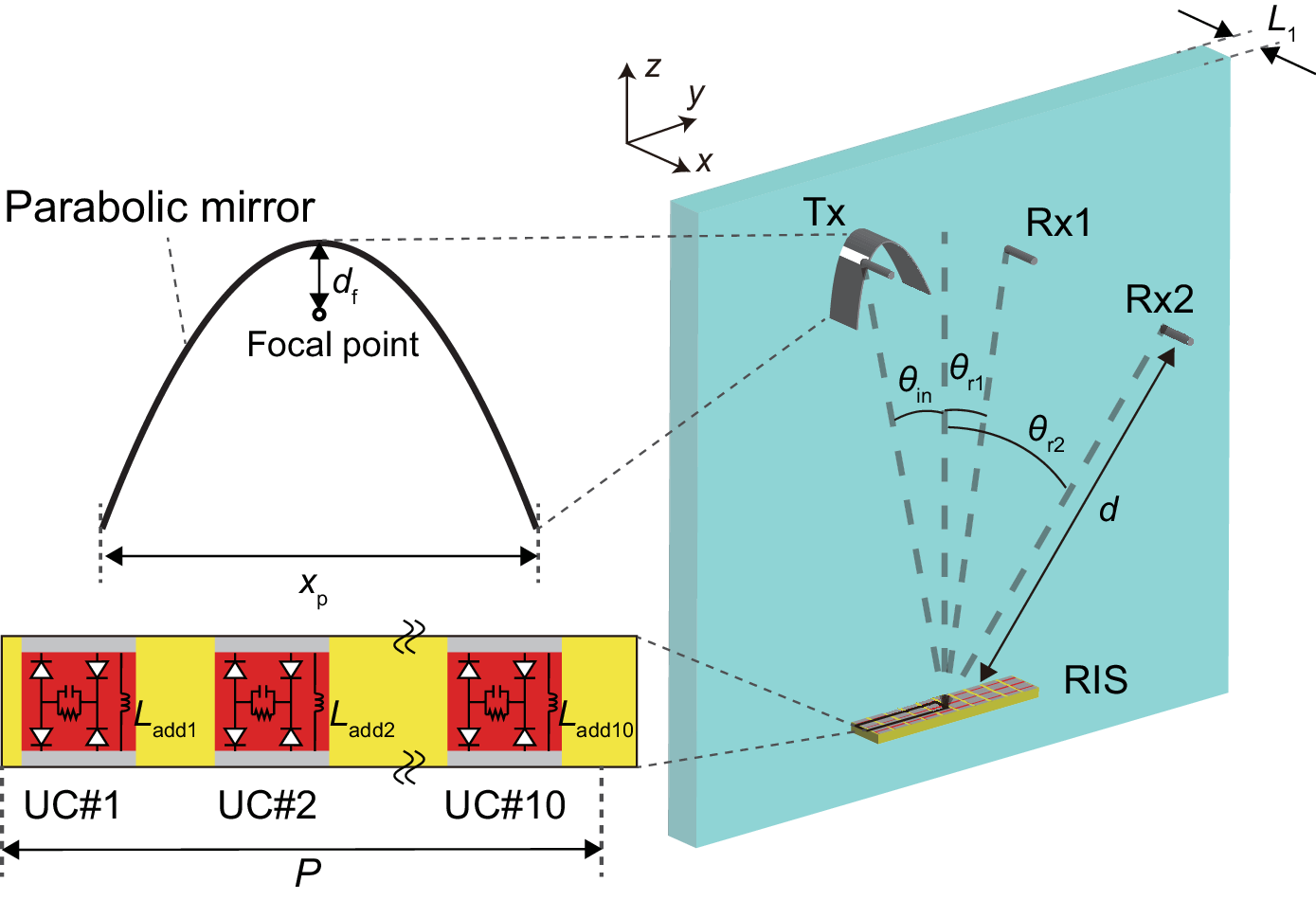}
    \caption{A rigorous RIS model including nonlinear circuit components in a compact simulation space. The RIS reflected the signals of Tx to two receivers, namely, Rx1 and Rx2, with a reduced far-field distance due to the presence of the parabolic mirror. The geometrical parameters adopted in this model are provided in Table \ref{tab:7}.}
    \label{fig:5}
\end{figure}

Following the acquisition of simulation results for the simplified RIS model in Fig.\ \ref{fig:4}(b), we built a rigorous RIS model that fully accounted for the fine-conducting geometries of UC$\#$1 to UC$\#$10, including loaded discrete circuit elements. To maintain a compact demonstration space even for a later experimental test, we modeled only one unit cell along the $x$ axis (the incident electric field ($E$ field) direction) of Fig.\ \ref{fig:5} between perfect electric conductor (PEC) boundaries. Therefore, the periodicity of the unit cells was ensured along the $x$ axis in this configuration by deploying RIS unit cells along the $y$ axis. The far-field distance was also reduced by surrounding a grounded 25-mm-tall monopole antenna Tx with a parabolic mirror, which contributed to realizing a plane wavefront with a short distance at the RIS location (see the design of the parabolic mirror in Fig.\ \ref{fig:5}). In contrast, two receivers, Rx1 and Rx2, were not surrounded by parabolic mirrors to avoid strong interactions with the reflected wave. Additionally, unlike the simulation model of Fig.\ \ref{fig:4}(a), where the incident angle was set to 0 degrees, the simulation model of Fig.\ \ref{fig:5} involved setting the incident angle $\theta _{in}$ to 10 degrees to suppress the influence of a standing wave for a specular reflection. The geometric parameters of the rigorous RIS model are shown in Table \ref{tab:7}. 

\begin{table}
\caption{The geometric parameters adopted in the rigorous RIS simulation model of Fig.\ \ref{fig:5}. \label{tab:7}}
\renewcommand{\arraystretch}{1.2}
\centering
\begin{tabular}{|c| c| c|}
\hline
Parameter & Unit                  & Value \\ \hline\hline
$\theta_{in}$        & degree                     & 10                              \\ 
$\theta_{r1}$        & degree                     & 10                              \\ 
$\theta_{r2}$        & degree                     & 50                              \\ 
$d$      & mm                    & 1100                                         \\ 
$d_f$      & mm                    & 20                                         \\ 
$x_p$      & mm                    & 160                                         \\ 
$L_1$      & mm                    & 34                                         \\ 
\hline
\end{tabular}
\end{table}

\begin{figure}
\subfigure[]{
\centering
\includegraphics[width=\linewidth]{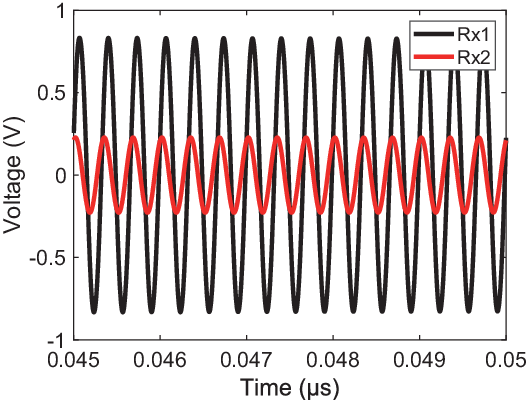}
}
\subfigure[]{
\centering
\includegraphics[width=\linewidth]{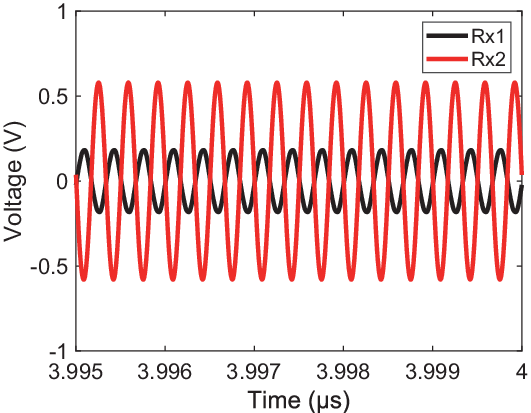}
}
\caption{The voltages received by Rx1 and Rx2 in the rigorous RIS model of Fig.\ \ref{fig:5}. (a) The SP scenario and (b) the CW scenario.}
\label{fig:6}
\end{figure}

\begin{figure}
\subfigure[]{
\centering
\includegraphics[width=\linewidth]{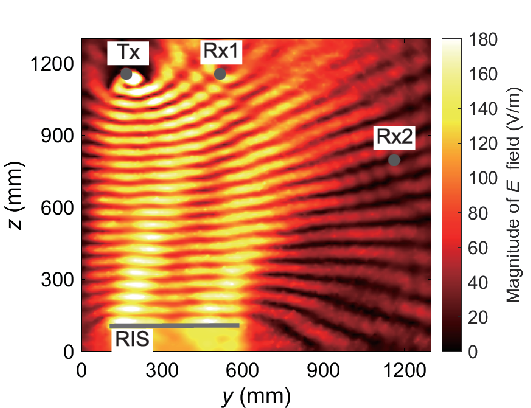}
}
\subfigure[]{
\centering
\includegraphics[width=\linewidth]{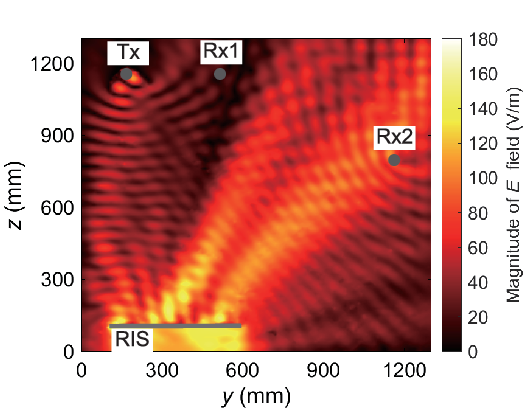}
}
\caption{The $E$ field distributions of the rigorous RIS model shown in Fig.\ \ref{fig:5}. (a) The SP scenario and (b) the CW scenario.}
\label{fig:7}
\end{figure}

When the incident frequency and power were set to 3.00 GHz and 30 dBm, respectively, receiver signals were obtained during the initial period and in the steady state, as shown in Fig.\ \ref{fig:6}(a) and Fig.\ \ref{fig:6}(b), respectively. First, when an SP was incident on the RIS at $\theta_{in}=10^\circ$, the incident wave was more strongly received at $\theta_{r1}$ (= $10^\circ$) than at $\theta_{r2}$ (= $50^\circ$) as the zeroth order or a specular reflection, which is represented in Fig.\ \ref{fig:6}(a) by the Rx1 voltage being larger than the Rx2 voltage. In contrast, for the CW (i.e., the steady-state) scenario, the resonance condition self-adjusted from specular reflection to a phase-modulating state, resulting in a high reflection power (or voltage) at the anomalous angle of $\theta_{r2}$ (at Rx2 in Fig.\ \ref{fig:6}(b)). The change exhibited by the reflecting state is more clearly demonstrated in Fig.\ \ref{fig:7}, where the scattered $E$ field profiles are visualized over a two-dimensional surface. For example, Fig.\ \ref{fig:7}(a) indicates that the $E$ field reflected by our RIS was mostly directed toward Rx1. Therefore, a larger magnitude was obtained for the $E$ field around Rx1. However, the magnitude of the $E$ field was lower for the CW case of Fig.\ \ref{fig:7}(b), where the RIS alternatively directed the reflected wave toward Rx2. The results in Figs.\ \ref{fig:6} and \ref{fig:7} ensure that our passive RIS exhibited two beam profiles with a maximum amplitude shifting between 10$^\circ$ during pulse illumination and 50$^\circ$ during CW illumination. This functionality was achieved through the abovementioned capacitor-based waveform-selective metasurfaces, which were characterized by pulse width-dependent reflection \cite{wakatsuchi2019waveform}. We note that this pulse width-dependent reflection scheme can readily be tailored by replacing the capacitor-based circuit with a different configuration, such as an inductor-based circuit and a combination of these two circuits \cite{wakatsuchi2015waveform}. In practice, our metasurface inherently possesses three distinct states, including the absorptive state \cite{qu2015tailor}. However, for our present configuration, we did not exploit the absorptive state that might appear when intermediate pulses interact with a metasurface, as reported in our earlier work \cite{wakatsuchi2015waveform}. 

\section{Experimental Validation of the RIS}

\begin{figure}[b!]
    \centering
    \includegraphics[width=\linewidth]{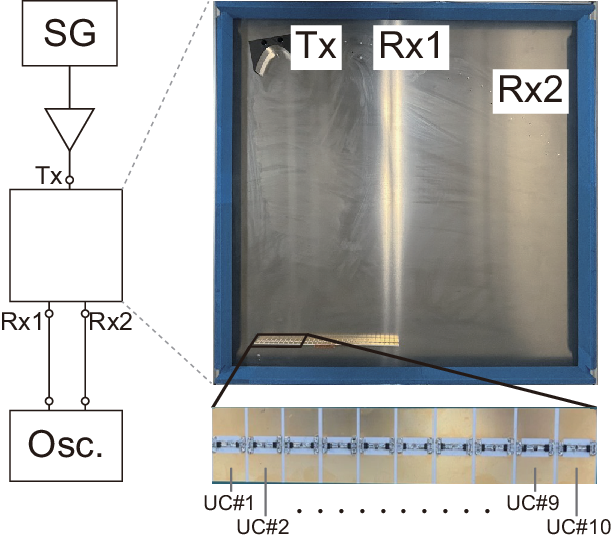}
    \caption{The measurement setup, including the RIS prototype. }
    \label{fig:8}
\end{figure}

Next, we experimentally evaluated the reflecting profile of the proposed RIS. For the experimental validation, we adopted the same conditions as those used by the simulation model of Fig.\ \ref{fig:5}. For example, we used a parallel-plate waveguide and absorbing forms to have dimensions that were identical to those of the simulation setup, as shown in Fig.\ \ref{fig:8}. In the measurement setup, a signal generator (Anritsu, MG3690C) was used to generate incident signals that were amplified by an amplifier (Ophir RF, 5193) to increase the signal magnitude to 30 dBm at 3.52 GHz and turn on the diodes integrated with our RIS. The incident signals were then radiated from a grounded monopole antenna surrounded by a parabolic mirror composed of a stainless copper plate and a supporting resin. Therefore, the incident signals were directed to a prototype of the RIS that was expected to vary its reflecting characteristics in accordance with the incident pulse duration. The RIS was also fabricated by following the physical dimensions and circuit values that were applied to the simulation model. Rx1 and Rx2 were connected to an oscilloscope (Teledyne Lecroy, 9404M) via coaxial cables so that the received signals were observed in the time domain to obtain their corresponding reflectance profiles. 

\begin{figure}
\subfigure[]{
\centering
\includegraphics[width=\linewidth]{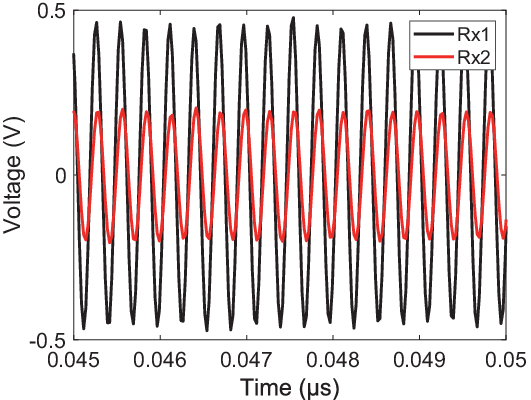}
}
\subfigure[]{
\centering
\includegraphics[width=\linewidth]{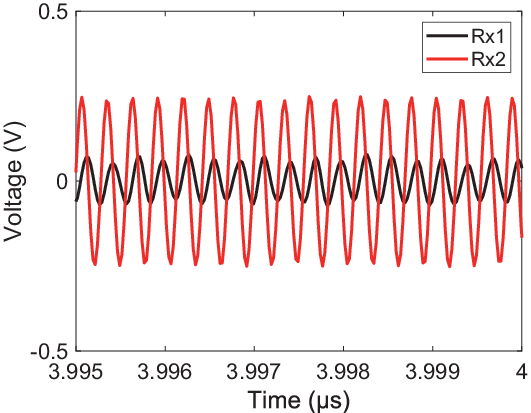}
}
\caption{The measurement results obtained for the received signals at Rx1 and Rx2. (a) SP conditions and (b) CW conditions.}
\label{fig:9}
\end{figure}

The measurement results obtained for the signals received at Rx1 and Rx2 are shown in Fig.\ \ref{fig:9}(a) and Fig.\ \ref{fig:9}(b) for the SP and CW conditions, respectively. Although the measured data were smaller than the simulation results shown in Fig.\ \ref{fig:6}, a similar conclusion was drawn from the measurements of Fig.\ \ref{fig:9}. Specifically, the signal magnitude of Rx1 was larger than that of Rx2 in the SP case because of the specular reflection state of the RIS. However, a completely different trend was demonstrated in the CW case, where the input signal was more strongly received by Rx2 than by Rx1 as the RIS switched to the anomalous reflection state. Thus, these measurement results also support the passive switchable beamforming performance of our RIS without any external active control system or DC voltage supply, unlike in conventional studies. 

\section{Communication Performance of the RIS}
\begin{figure}
\subfigure[]{
\centering
\includegraphics[width=\linewidth]{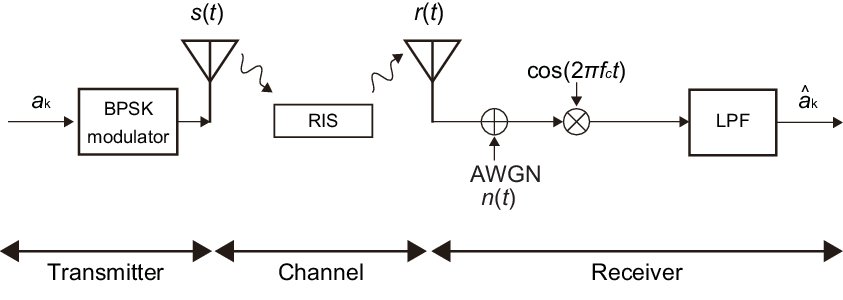}
}
\subfigure[]{
\centering
\includegraphics[width=\linewidth]{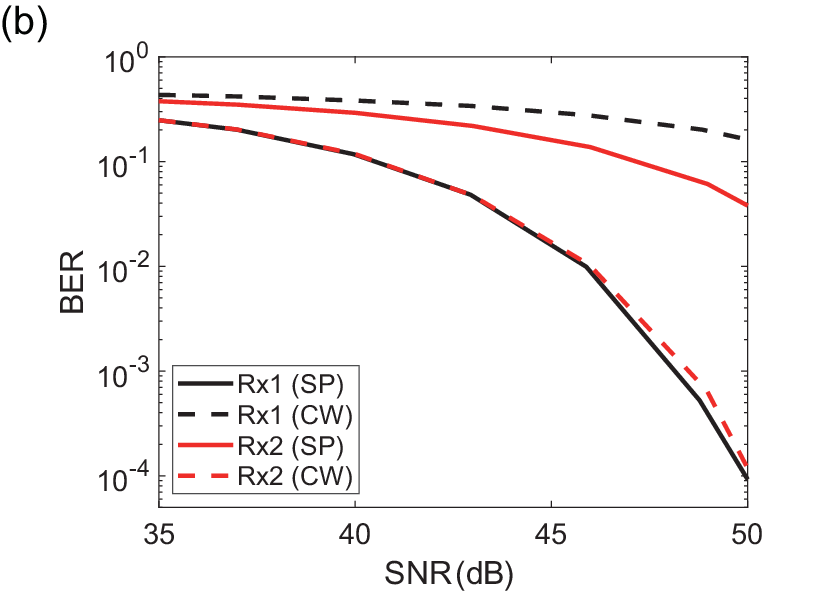}
}
\caption{Numerical evaluation of the achieved communication performance. (a) Block diagram of the BPSK modulation process. (b) The BERs of Rx1 and Rx2 under SP and CW conditions.}
\label{fig:10}
\end{figure}

Moreover, the communication performance of the developed RIS was assessed by using our simulation model and an experimental prototype in a practical scenario that included a modulation scheme. Under this circumstance, the RIS conducted variable beamforming and thereby adjusted the signal strengths at both receivers. Here, we employed the binary phase shift keying (BPSK) scheme to modulate both the SP and CW carrier signals, for which the RIS could respond differently. Fig.\ \ref{fig:10}(a) depicts the block diagram of the system model, with $a_k$ representing the input binary data (i.e., 0 or 1) and $f_c$ denoting the frequency of the carrier signal (specifically, 3 GHz). Each bit underwent phase modulation via a cosine function, which was characterized by a 10-ns signal duration and a binary phase condition of $\pm{90^\circ}$ as follows:
\begin{eqnarray}
   s(t)=A\cos{(2\pi f_c t+a_k \pi)},
   \label{eq:9}
\end{eqnarray}
where $s(t)$ and $A$ denote the time-domain waveform and amplitude of the input signal, respectively, and $t$ represents time. At the receiver end, the received signals passed through an additive white Gaussian noise (AWGN) component $n(t)$, which was followed by multiplication of the carrier signal and a subsequent harmonics elimination process implemented via a low-pass filter (LPF). Accordingly, the bit error rate (BER) was assessed at the distinctive locations of the two receivers (i.e., Rx1 and Rx2) by extracting a single point discretized in the time domain of each signal. 

First, for a numerical examination of the pulse width-dependent beamforming capabilities of the RIS, we simulated the system dynamics using the previously demonstrated electromagnetic (EM) model to depict the wireless link channel. In alignment with the methodology described in Section \ref{sec:sim}, we adopted a communication setup incorporating two monopole receivers and one monopole transmitter surrounded by a parabolic mirror, as illustrated in Fig.\ \ref{fig:5}. To implement BPSK-modulated signals, the EM model subsequently interfacing with the ANSYS circuit solver was configured to realize transient analysis. The time-varying signals derived from the simulation were demodulated using MATLAB. 

The simulated BERs are shown in Fig.\ \ref{fig:10}(b), where first, the BER of Rx1 during the initial state was compared with that of Rx2 in the steady state. In these scenarios, the received signal magnitudes were larger than those of the other scenarios in Figs. \ref{fig:6} and \ref{fig:7} (i.e., the received signal of Rx1 in the steady state and that of Rx2 during the initial state). Additionally, in contrast with Figs. \ref{fig:6} and \ref{fig:7}, where the bandwidths of the input signals were almost identical to the oscillation frequency \cite{wakatsuchi2019waveform, wakatsuchi2015time}, the signals adopted in Fig.\ \ref{fig:10} were modulated as BPSK signals and thus had wider bandwidths. Nonetheless, these results indicate that the BER of Rx1 during the initial state and that of Rx2 in the steady state remained relatively low, which was more clearly understood by plotting the BERs of the other scenarios, namely, Rx1 in the steady state and Rx2 during the initial state. Compared with the performances achieved in the latter two scenarios, the BERs of the former two cases improved by more than 7 dB. 

\begin{figure}
\subfigure[]{
\centering
\includegraphics[width=\linewidth]{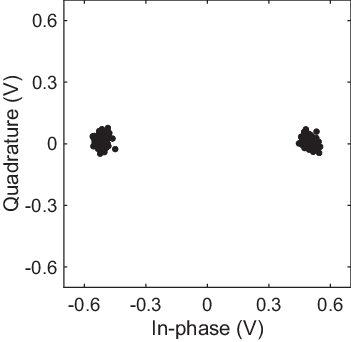}
}
\subfigure[]{
\centering
\includegraphics[width=\linewidth]{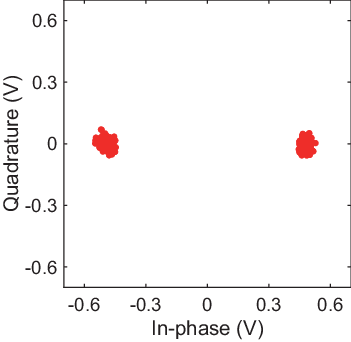}
}
\caption{Measurement results obtained for BPSK constellations. (a) Results of Rx1 under SP conditions. (b) Results of Rx2 under CW conditions.}
\label{fig:11}
\end{figure}

Second, for an experimental validation, a comparable measurement setup was employed to evaluate the communication performance of the RIS via constellation diagrams. Although the measurement setup was based on that adopted in Fig.\ \ref{fig:8}, the signal source was replaced with an arbitrary waveform generator (AWG) (Keysight Technologies, M8195A) to design more complicated modulated signal waveforms instead of sinusoidal waveforms. Under this circumstance, constellation diagrams were observed, as shown in Fig.\ \ref{fig:11}, which clearly indicates that the received signals were obtained around the zero- or one-$\pi$ location, as expected. Thus, these experimental results of the constellation diagrams, together with the numerical validation of the BERs in Fig.\ \ref{fig:10}, support the notion that the proposed RIS design concept is capable of leveraging the pulse width dependence even at a fixed frequency (or within a fixed frequency band), thereby controlling not only electromagnetic scattering profiles but also communication characteristics without any external DC supply, unlike the existing RIS approaches. 

\section{Discussion}
In this study, we reported the performance of an RIS design based on pulse width-dependent subwavelength structures, which are referred to as waveform-selective metasurfaces. The proposed RIS altered its electromagnetic responses and reflecting characteristics in accordance with the incident pulse width at a constant frequency. Unlike the conventional approaches based on active control systems, the proposed RIS was free of an external DC supply and control lines. Also, our RIS did not require symbol-level synchronization between the BSs and receivers and simplified the wireless communication systems. Moreover, the proposed design concept achieved dual reflection angles and suppressed the sidelobes of the spatial reflecting profiles, which was validated experimentally for the first time. However, the proof-of-concept RIS design can be further improved to attain more powerful and effective performance in wireless communication environments. First, the proposed RIS design was validated in a compact space bound between conducting parallel plates, which indicated that the reflecting beam angle was steered along only one direction. Additionally, we designed a linear phase gradient pattern over the RIS to tilt the reflected wavefront for the sake of simplicity. However, receivers can more efficiently collect communication signals if the reflected signal is focused on a smaller spot \cite{pfeiffer2013metamaterial, fathnan2022method}. Moreover, the power level of the input signals used in this study was much higher than that of ordinary communication signals since we adopted commercial diodes that turned on at a large voltage level, which can possibly be lowered by exploiting the latest design technologies related to semiconductors \cite{tanikawa2020metasurface}. However, if these issues are adequately addressed, the proposed RIS design can work more powerfully in wireless communication environments to compensate for the drawbacks of conventional active RISs, together with other pulse width-dependent technologies proposed in the domains of antenna and sensor design \cite{vellucci2019waveform, ushikoshi2023pulse, tachi2025multipath}, signal processing \cite{f2020temporal}, electromagnetic compatibility \cite{wakatsuchi2019waveform}, and frequency assignment \cite{takeshita2024frequency}. Finally, we note that our RIS design can be customized to achieve a higher degree of freedom for wireless communications by exploiting, for instance, polarization dependence \cite{sugiura2021joint}, orbital angular momentum (OAM) \cite{lloyd2017electron}, and other modulation schemes \cite{goldsmith2005wireless, hanzo2004quadrature}. 

\section{Conclusion}
The proposed concept demonstrates the ability to control the beamforming process using an RIS that switches between two singular reflection angles in response to variations exhibited by the pulse widths of incoming wireless signals. This RIS design offers a promising solution to the synchronization challenges that are encountered by the existing wireless communication systems while requiring only straightforward (DC voltage supply-free) hardware and a basic control algorithm that is based on controlling the pulse widths of the carrier signals. Structurally, the RIS integrates a waveform-selective metasurface comprising metallic patches that are embedded with transient and nonlinear circuits atop a grounded dielectric layer. This RIS strongly reflects CW signals with a phase gradient spanning 2$\pi$ within a single supercell period. In contrast, the RIS still maintains the reflectance magnitude of SP signals but with negligible phase variation within the same single period. Thus, two independent beam profiles can be obtained using either SP or CW signals. To validate the performance of the developed approach in wireless communication scenarios, we constructed and modeled the proposed RIS system incorporating BPSK modulation and linking the RIS-managed wireless link between a singular transmitter and dual receivers. Through numerical simulations and experiments, we effectively validated the proposed beamforming principle on the basis of pulse width variations. Our results may present new opportunities for simplifying wireless communication networks without active control systems and exploring pulse width-dependent RISs in practical scenarios involving next-generation wireless communication environments. 

\ifCLASSOPTIONcaptionsoff
  \newpage
\fi

\bibliographystyle{IEEEtran}

\vfill

\end{document}